# Structural, optical, magnetic and electrical properties of $Zn_{1-x}Co_xO$ thin films

**M Tay[1,2], Y H Wu[1*], G C Han[2], Y B Chen[3], X Q Pan[3], S J Wang[4], P Yang[5] and Y P Feng[6]**

[1] Department of Electrical and Computer Engineering, National University of Singapore, 4 Engineering Drive 3, Singapore 117576

[2] Data Storage Institute, 5 Engineering Drive 1, Singapore 117608

[3] Department of Materials Science and Engineering, University of Michigan, Ann Arbor, Michigan 48109-2136, USA

[4] Institute of Materials Research and Engineering (IMRE), 3 Research Link, Singapore 117602

[5] Singapore Synchrotron Light Source, Singapore 117603

[6] Department of Physics, National University of Singapore, Singapore 117576

---

[*] E-mail: elewuyh@nus.edu.sg

# Abstract

Despite a considerable effort aiming at elucidating the nature of ferromagnetism in ZnO-based magnetic semiconductor, its origin still remains debatable. Although the observation of above room temperature ferromagnetism has been reported frequently in literature by magnetometry measurement, so far there has been no report on correlated ferromagnetism in magnetic, optical and electrical measurements. In this paper, we investigate systematically the structural, optical, magnetic and electrical properties of $Zn_{1-x}Co_xO:Al$ thin films prepared by sputtering with x ranging from 0 to 0.33. We show that correlated ferromagnetism is present only in samples with x > 0.25. In contrast, samples with x < 0.2 exhibit weak ferromagnetism only in magnetometry measurement which is absent in optical and electrical measurements. We demonstrate, by systematic electrical transport studies that carrier localization indeed occurs below 20-50 K for samples with x < 0.2; however, this does not lead to the formation of ferromagnetic phase in these samples with an electron concentration in the range of $6 \times 10^{19}$ $cm^{-3}$ ~ $1 \times 10^{20}$ $cm^{-3}$. Detailed structural and optical transmission spectroscopy analyses revealed that the anomalous Hall effect observed in samples with x > 0.25 is due to the formation of secondary phases and Co clusters.

1       Introduction

Since the theoretical predication of room-temperature ferromagnetism in diluted magnetic semiconductors (DMSs) [1], much work, both theoretical and experimental, has been devoted to the investigation of ferromagnetism in ZnO doped with various transition metals [2]. Although most theoretical predications as well as majority of experimental studies favour room-temperature ferromagnetism in transition metal doped ZnO, the origin of observed ferromagnetic behaviour remains debatable, and in particular, the existence of carrier-mediated ferromagnetism has yet to be demonstrated. The mean-field Zener model of Dietl *et al.* [1] predicts a Curie temperature ($T_C$) above room temperature for ZnO doped with 5% Mn and with a hole concentration of $3.5 \times 10^{20}$ cm$^{-3}$. Using the first-principles calculation, Sato and Katayama-Yoshida [3] and Lee and Chang [4] predict that heavy electron doping and high Co composition are the key to obtaining ferromagnetic $Zn_{1-x}Co_xO$. On the other hand, Spaldin argues that only hole doping promotes ferromagnetism in both $Zn_{1-x}Co_xO$ and $Zn_{1-x}Mn_xO$ [5]. Very recently, Sluiter *et al.* predict that both hole doping and electron doping promote ferromagnetic ordering in $Zn_{1-x}Co_xO$ and $Zn_{1-x}Mn_xO$ [6]. Park and Chadi have shown that spin-spin interaction mediated by high-concentration of interstitial hydrogen can lead to high temperature ferromagnetism in $Zn_{1-x}Co_xO$ in the absence of carriers [7]. Coey *et al.* argue that conventional superexchange or double-exchange interactions cannot produce long-range magnetic order at low concentrations of magnetic doping and propose that ferromagnetism in oxides is mediated by a donor impurity band which splits globally when bound magnetic polarons (BMPs) merge [8]. In highly concentrated magnetic semiconductors, Sato *et al.* showed that spinodal decomposition inherently occurs due to strong attractive interactions between impurities, leading to high Curie temperature ferromagnetism due to the formation of magnetic networks [9]. As is with the case of theoretical studies, intensive experimental investigations so far have also only produced widely diverging results in Co-doped ZnO, ranging from intrinsic ferromagnetism with various Curie temperatures [10-17] to ferromagnetism with extrinsic origins [18-20], paramagnetism or superparamagnetism [21-23] and anti-ferromagnetism [24-26]. In addition to the high-sensitivity of magnetic properties to preparation techniques and conditions, the rather chaotic situation is also caused by the lack of a commonly agreeable way to determine if a DMS is intrinsic when a magnetic moment versus applied field curve with a very small hysteresis is observed in a simple magnetometry measurement. The common approach taken so far by most experimental work which supports the existence of intrinsic ferromagnetism in $Zn_{1-x}Co_xO$ is as follows. First, the existence of ferromagnetism is "confirmed" by either direct measurement using a magnetometer or the combination of this with indirect measurements such as magnetic circular dichroism (MCD) and x-ray MCD (XMCD). Second, the presence of precipitates and other secondary phases is "excluded" by high-resolution transmission electron microscopy (HRTEM) and x-ray diffraction (XRD) measurements. Third, substitution of Zn with Co into the host matrix is confirmed by optical absorption, magnetooptical spectroscopy (including the confirmation of coupling between band carriers with d electrons of the Co ions) or other valence analysis techniques. We argue that the validity of this approach is only relevant if all the measurements are performed on the same physical location of the same sample as well as if the measured results are well correlated with each other. In the latter case, in addition to magnetic and magnetooptical measurements, it is also of great importance to perform systematic studies of electrical transport properties. Given the fact that it is almost impossible to ensure that all the sample preparation techniques and conditions are identical for experiments performed at different groups, researchers should at least make efforts to conduct systematic studies using their own specific experimental setups. In order to draw meaningful conclusions from the experiments, it is crucial that all different types of characterizations should be carried out on same samples instead of samples with similar chemical concentrations.

In a recent paper, we have studied the structural, optical, magnetic and electrical transport properties of both co-doped and δ-doped ZnO:Co thin films and have reported that the correlated ferromagnetism in both types of samples was due to extrinsic origin [27]. In this paper, we report on a systematic study of $Zn_{1-x}Co_xO$ thin films by focusing on co-doped samples only but by varying the Co composition more systematically in a much smaller step from x = 0 to 0.33. All samples have been characterized thoroughly using various techniques. Special attentions have been paid to ensure that a series of characterization experiments could be performed on each of the series of samples studied. Although all samples were found to contain ferromagnetic phases up to room temperature in measurements by a commercial superconducting quantum interference device (SQUID) magnetometer, it was found that only samples with x > 0.25 exhibit correlated ferromagnetism in their optical, electrical and magnetic properties measured by MCD, anomalous Hall effect (AHE) and SQUID, respectively. Detailed structural and optical transmission spectroscopy analyses revealed that ferromagnetism observed in samples with x > 0.25 is due to formation of secondary phases and Co clusters. In contrast, samples with x < 0.2 "appears" to be ferromagnetic embedded within a paramagnetic background in SQUID measurement, but there is no AHE observed even though their magnetotransport properties are dominated by sd interactions.

## 2    Experimental

The $Zn_{1-x}Co_xO$ (x = 0–0.33) thin films were deposited on (001) sapphire (α-$Al_2O_3$) substrates in a high vacuum chamber with a base pressure < $1 \times 10^{-7}$ Torr using a combination of radio-frequency (RF) and direct current (DC) magnetron sputtering. Sintered ZnO, $Al_2O_3$ and Co materials were used as the sputtering sources for ZnO, Al and Co, respectively. All the samples were sputtered in an atmosphere of pure Ar gas at a pressure of 5 mTorr. Prior to the deposition, the substrates were cleaned using Ar reverse sputtering at 20 mTorr in the pre-cleaning chamber. Films were grown at a substrate temperature of 500 °C, ZnO sputtering power of 150 W and Al$_2$O$_3$ sputtering power of 30 W. Growth conditions have been optimized to produce Al-doped ZnO films with a resistivity of about 1.3 mΩ.cm (Al composition < 0.1%). The Co composition is varied via Co sputtering power (3-50 W). The actual Co compositions were measured by x-ray photoelectron spectroscopy (XPS). In our specific setup, the minimum controllable power is about 3 W which gives a Co composition of about 5 at.% as determined by XPS. The Co composition of the other samples deposited by varying the Co sputtering power from 8 W to 50 W is listed in Table 1. The Co composition increases monotonically with the Co sputtering power below 35 W, above which it is almost constant. The microstructures of these samples were characterized using XRD, high-angle angular dark field (HAADF), high resolution transmission electron microscopy (HRTEM) and electron energy loss spectroscopy (EELS). The optical and magnetooptical properties of the samples were characterized using a UV-visible spectrophotometer and by MCD measurement, respectively. The magnetic properties of the sample were characterized using a commercial SQUID magnetometer. For electrical characterization, Hall bars with a length of 324 μm and a width of 80 μm were fabricated for each sample using a direct laser writer. A 1-3-3-1 eight-contact Hall bar configuration was used to measure both the longitudinal and Hall voltages. The same sample geometry has also been used to measure the magnetoresistance. The thickness of all the films prepared is around 200 nm as confirmed by the TEM observation.

## 3  Results and Discussion

### 3.1  Structural properties

Figures 1(a)-(c) show the XRD patterns of Zn$_{1-x}$Co$_x$O:Al with different Co compositions. The data are displayed for three different ranges of x values, i.e., (a) $x \leq 0.2$, (b) $0.2 < x < 0.3$ and (c) $x \geq 0.3$. For samples with $x \leq 0.2$, the XRD patterns in the range of $2\theta = 30\text{-}50°$ consist of only the Zn$_{1-x}$Co$_x$O (002) peak and peaks associated with the sapphire substrate (including the two sharp peaks at 2θ~ 40.5° and ~ 42.3°). As the ionic radius of $Co^{2+}$ is about 96% of that of $Zn^{2+}$, the in-plane lattice constant of relaxed $Zn_{1-x}Co_xO$ film is expected to decrease when Zn atoms are replaced by Co atoms, leading to an increase of out-of-plane lattice constant due its large Poisson's ratio [28]. This explains why the (002) peak of $Zn_{1-x}Co_xO$ shifts to the lower angle side of the original (002) peak of ZnO, as shown in Fig. 1(a). This is also an indication that within this composition range, the Co atoms are soluble in ZnO [29,30]. For samples with x > 0.2, as shown in Fig. 1(b), three new peaks appear at 2θ ≈ 31.8°, 35.8°-36° and 44.5°, respectively. The assignment of these peaks is nontrivial because Co may exist in the material in question in at least five different forms: $Zn_{1-x}Co_xO$, CoO, $Co_3O_4$, $ZnCo_2O_4$ and Co. Furthermore, Co and CoO nanoparticles may exist in both cubic and hexagonal structures [31]. For instance, the peak near 36° may be assigned to either one of the following peaks: $Zn_{1-x}Co_xO$ (101), CoO (111) at 36.493° for cubic CoO, CoO (101) at 36.3° for hexagonal CoO, $ZnCo_2O_4$ (311) at 36.803° and $Co_3O_4$ (311) at 36.853°. However, considering the fact that bulk ZnO (101) peak is at 36.253° and after doping it may shift to lower angle just like the (002) peak, the peak near 36° may be assigned to the $Zn_{1-x}Co_xO$ (101) diffraction. Similarly, the peak at 31.8° is due to $Zn_{1-x}Co_xO$ (100). This indicates that polycrystalline $Zn_{1-x}Co_xO$ forms in the Co composition range of 0.2 < x < 0.3, though below x = 0.2 the film is mainly (002) textured. As we will discuss shortly, the EELS analysis showed that the valence of Co in samples with x <0.3 is dominantly 2+, suggesting that the main phase is $Zn_{1-x}Co_xO$. On the other hand, the peak at 44.5° is near peak positions of Co (111) at 44.217°, $ZnCo_2O_4$ (400) at 44.74° and $Co_3O_4$ (400) at 44.81°. As can be seen from Fig. 1(c), peak intensity at 31.8° and 36.1° decreases, while those near 44.5° increases when x exceeds 0.3. This trend strongly suggests that the peak at 44.5° is due to Co clusters, though again we cannot exclude the existence of other secondary phases such as $ZnCo_2O_4$ and $Co_3O_4$. The former is more likely because the formation of $ZnCo_2O_4$ and $Co_3O_4$ needs an oxygen-rich environment instead of more Co atoms. We also noticed that the ZnO (002) peak shifts back

to the original position when x exceeds 0.3; this is attributed to the phase segregation of CoO from the host matrix, which in turn leads to the relaxation of strains. Before we end this session, we have to point out that the two very sharp peaks, one at $2\theta = 40.6^0$ and the other at $2\theta = 42.26^0$ are due to the diffraction from substrate, presumably from some high order planes. They only appear when the measurement is done using a synchrotron light source.

Figures 2(a)–(c) show the low magnification transmission electron microscope (TEM) image of samples D (x = 0.2), E (x = 0.24) and H (x = 0.29), respectively. Insets are the corresponding diffraction patterns. Although it is not shown here, we have also carried out HAADF analysis and energy dispersive spectroscopy (EDS) mapping of these samples. For sample D, both EDS mappings & HAADF images showed that there is no obvious Co-segregation and precipitation in the ZnO host matrix. Diffraction patterns showed that the film is in a single crystalline phase. Similarly, for sample E, there was no obvious Co-segregation and precipitation observed in the EDS and HAADF results, and the diffraction patterns showed that the film is grown epitaxially on the substrate. However, some additional spots are observed in the diffraction patterns of which the origin is not clear. In a sharp contrast to samples D and E, samples H exhibits a columnar growth mode, similar to that reported by Schaedler *et. al.* [30]. The electron diffraction analysis (Fig. 2(d)) shows that the inhomogeneous sample contains secondary phases ($ZnCo_2O_4$ and CoO) and Co clusters (see HRTEM image in Fig. 2(e)), The EELS spectra taken at four randomly chosen locations of the film all showed that the valence state of Co is 2+. Therefore, the film of sample H consists of mainly $Zn_{1-x}Co_xO$ and CoO because the valence state of Co in $ZnCo_2O_4$ and $Co_3O_4$ is 3+ and 4+, and that in Co clusters is 0, respectively. As the lattice constant of hexagonal CoO is very close to that of ZnO [31]; it is very difficult to differentiate CoO from ZnO by XRD or TEM observation if the former grows pseudmorphically inside the ZnO host matrix. However, as we will discuss below in the optical transmission measurement, the $Zn_{1-x}Co_xO$ films with x < 0.24 contain both Zn-rich and Co-rich

regions. The former shows a bandgap larger than that of ZnO, while the latter has a bandgap which is very close to that of CoO.

3.2 Optical properties

One of the common ways used to determine if Co atoms have replaced Zn to form substitutional dopants is to observe if there are clear d-d transition lines in the optical transmission spectrum. Figure 3(a) shows the transmittance of samples with different Co compositions normalized to their respective values at 800 nm. In addition to band edge absorptions, absorption bands were also observed at 571 nm, 618 nm, and 665 nm which are attributed to d-d transitions of tetrahedrally coordinated $Co^{2+}$. They are assigned as $^4A_2(F) \rightarrow {}^2A_1(G)$, $^4A_2(F) \rightarrow {}^4T_1(P)$, and $^4A_2(F) \rightarrow {}^2E(G)$ transitions in high spin state $Co^{2+}(d^7)$, respectively [32]. With the increase of Co composition, the absolute strength of these absorption bands increases almost linearly, whereas the amplitude of the absorption fringes initially increases and then decreases. The former suggests that majority of Co atoms substitute Zn to form $Co^{2+}(d^7)$ ions, reducing the possibility of Co cluster formation. The latter can be ascribed to the fluctuations in local crystal field surrounding different Co ions, in particular due to the formation of Co-Co bonds and Co clusters at very high Co compositions. To probe further into the details of differences among different samples, in particular at the band edge region, we differentiate the transmittance (T) with respect to the wavelength (λ) and show the results in Fig. 3(b) (dT/dλ – λ). It is apparent that peaks in Fig. 3(b) are corresponding to the maximum rate of change of transmittance with respect to the wavelength rather than transition energies associated with the absorption of photons. For example, in the case of Al-doped ZnO without Co doping (x = 0), a strong peak appears at 345.5 nm which is much shorter than the wavelength corresponding to ZnO bandgap (370 nm). Nevertheless, the differentiation does reveal clearly the presence of different absorption bands. As is with the case of XRD data, optical transmission spectra can also be divided into 3 groups based on their x values. In the first group for which x ≤ 0.2 (sample A-D), there is a clear blueshift of the band edge absorption as compared to Al-doped ZnO without Co doping [33]. This agrees well with XRD results that in this composition range Co atoms are soluble in ZnO and form Co-incorporated ZnO phase [30]. In addition to this blue shifted peak (A), there is also a red-shifted peak (B). As it is shown in the inset of Fig. 3(b), the wavelength of peak A decreases, while that of peak B increases with increasing Co composition. However, the change is very small from x = 0 to 0.24. As x increases to the range between 0.24 and 0.29 (samples E-H), peak A disappears suddenly, and at the meantime, the wavelength of peak B increases sharply. This trend continues when the Co composition increases further to x > 0.3 (sample J and L). The energy differences between peak A and peak B are 0.41, 0.76, 0.77 and 0.75 eV for samples A, B, C and E, respectively. Kittilstved *et al.* observed an absorption band in $Zn_{1-x}Co_xO$ (x = 0.035) at an energy of 0.32 eV below the excitonic transition line of ZnO and assigned it to ligand valence band to metal charge transfer transitions [34]. It is not clear, however, how this energy level will vary with Co composition because only data for x = 0.035 was discussed.

Qiu *et al.* have observed an abnormal bandgap narrowing in $Zn_{1-x}Co_xO$ nanorods with x = 0 ~ 0.1 and attributed it to lattice volume expansion of ZnO induced by Co-doping. The redshift was found to follow the relationship $\Delta E_g = 0.54(e^{-x/0.03} - 1)$ eV [35]. As shown by the solid-line in inset of Fig. 3(b), the wavelength of peak B follows well this relationship below x = 0.24, i.e., $\lambda_B = 1239/\left[3.589 + 0.54(e^{-x/0.03} - 1)\right]$. On the other hand, the wavelength of peak A can be fitted well by the equation $\lambda_A = 1239/(3.589 + 1.69x)$ and the result is shown by the lower solid curve in the inset of Fig. 3(b). Here, the coefficient of x (1.69 eV) was taken from the fitting result of excitonic transitions in lightly doped samples discussed in reference [33]. By combining the XRD, EELS and optical transmission data, we argue that peak A shown in the inset of Fig. 3(b) is due to Zn-rich $Zn_{1-x}Co_xO$ phase, while peak B is due to Co-rich $Zn_{1-x}Co_xO$ phase.

Considering the facts that the XRD diffraction peaks are close to those of ZnO or hexagonal CoO, the wavelength of peak B is close to the bandgap of CoO [36] (~ 2.9 eV), and the valence of Co is dominantly 2+ below x = 0.29, the Co-rich $Zn_{1-x}Co_xO$ phase is most likely Zn-incorporated CoO. With a further increase of Co, the Zn-rich phase gradually disappears; instead the Co-rich phase and Co clusters become dominant, as shown by the XRD patterns in Fig. 1(c).

3.3 Magnetic and magneto-optical properties

The successful incorporation of Co as substitutional impurity in ZnO for x < 0.2 itself, of course, does not necessarily lead to the formation of intrinsic DMS in this system. In order to achieve this goal, there must be a strong carrier-spin interaction between band electrons or holes with the spin of $Co^{2+}$ ions as well as ferromagnetic coupling between $Co^{2+}$ ions via the carriers. Although strong s,p-d interactions have already been confirmed from the study of field-dependent excitonic transitions [33] and MCD in the vicinity of the bandgap [33,37,38], there is still no direct evidence to show that there is a ferromagnetic interaction among the localized spins of $Co^{2+}$ in dilute $Zn_{1-x}Co_xO$ samples. Using magnetometry and electron paramagnetic resonance measurements in combination with crystal field theory, Sati *et al.* revealed that isolated $Co^{2+}$ ions in ZnO possess a strong single ion anisotropy of $DS_z^2$ type, with D = 2.76 cm$^{-1}$, leading to a paramagnetic behaviour of dilute $Zn_{1-x}Co_xO$ [39]. It is argued that an "easy plane" ferromagnet could be formed if there is a ferromagnetic coupling between $Co^{2+}$ ions. Theoretically, ferromagnetic coupling may originate from carrier-mediated mechanism [1], spin-split donor impurity band [8] and other similar mechanisms [34]; however, their existence in dilute $Zn_{1-x}Co_xO$ samples has yet to be confirmed through the observation of correlated ferromagnetism in magnetic, magneto-optical and electrical measurements.

In more concentrated samples, the broadening of excitonic transitions makes it difficult to study the s,p-d interaction through giant Zeeman splittings; in these cases, MCD has proven to be a convenient technique to study the magnetic properties of transition metal ions in ZnO [37,38].

The main advantage of MCD is that it can detect signal from different phases owing to its energy selectivity. Figures 3(c) and 3(d) show the MCD hysteresis curves of sample H (x = 0.29) obtained at different photon energies at 6 K and 300 K, respectively (the energy positions are marked in the transmission spectrum shown in Fig. 3(a) by the "*" symbol). The MCD curves are strongly dependent on the photon energy, confirming again that the sample is inhomogeneous and consists of ferromagnetic regions of different phases. On the other hand, MCD curves of sample B shows only paramagnetic behaviour (not shown here). As it will be discussed later, only the MCD curve taken at 2.92 eV (425 nm) agrees well with the hysteresis curves measured by SQUID and AHE.

The M-H curves of samples A, B, D and H, measured at room temperature by SQUID, are shown in Fig. 4(a). Except for sample H, the M-H curves of samples A, B and D consist of a weak ferromagnetic and a strong paramagnetic phase (note that the ordinate is in logarithm scale). In order to focus on the ferromagnetic phase only, the paramagnetic contribution has been removed from the curves shown in Fig. 4(a). Although the ferromagnetic properties of samples A, B and D are weak, clear hysteresis has been observed. As the Co concentration is increased by 2-4 times from samples A to H, the magnetic moment remains to be small in samples A, B and D. However, a drastic increase in magnetic moment is observed from sample D to H (see Fig. 4(b)). On the other hand, as shown in Fig. 4(c), the in-plane coercivity initially increases with the Co composition and reaches a maximum at about x = 0.16; it starts to drop, reaching a minimum near x = 0.25, and increases again at about x = 0.27. Below x = 0.2, the films showed an in-plane anisotropy. The sudden drop of coercivity at x = 0.25 at high temperature is due to the switch over from in-plane to out-of-plane anisotropy caused by the appearance of secondary phases. The formation of Co clusters at x > 0.27 drives the coercivity to decrease again. The sharp increase in magnetic moment occurs in the region of x = 0.25-0.29, which agrees very well with the region where a sharp increase has been seen in the wavelength of peak B (inset of Fig. 3(b)). This suggests strongly that the sharp increase of magnetic moment originates from secondary phases. Bulk CoO in rock-salt structure is known to be an antiferromagnet with a Néel temperature of 297 K. However, small CoO nanoparticles can be ferromagnetic due to frustrated surface spins [40]. In addition to this, the incorporation of Zn or change of cubic structure to wurtzite structure [31] may also be responsible for the ferromagnetic properties. It is observed that sample H and other samples with x > 0.3 all exhibit a well defined perpendicular anisotropy (Fig. 4(e)) [41], whereas the M-H curves for samples with x < 0.2 shows a weak in-plane anisotropy in the paramagnetic phase (Fig. 4(d)). The latter agrees well with the observation of reference [39]. It should be noted that there is no direct evidence to show that the small hysteresis observed in samples A, B and D are due to $Co^{2+}$ ions because the SQUID simply picks up signals from all magnetic species. Therefore, it is not surprising to observe that sample B is ferromagnetic at room temperature in SQUID measurement, but paramagnetic in MCD measurement down to 6 K. Figure 4(f) compares the M-H curves for sample H obtained by different techniques. We will come back to discuss this figure shortly after the Hall measurement data are presented.

3.4     Electrical transport properties

3.4.1   Anomalous Hall effect

In addition to magnetometry and magneto-optical measurements, electrical transport measurements also play a crucial role in establishing the origin of magnetism in DMSs, in particular the Hall effect. The Hall resistivity, $\rho_{xy}$, in a ferromagnet is generally given by $\rho_{xy} = R_o B + R_s \mu_0 M$, where B is the magnetic induction, $\mu_0$ is the magnetic permeability in vacuum, M is the magnetization in field direction, $R_o$ is the ordinary Hall coefficient and $R_s$ is anomalous Hall coefficient [42]. The first term is due to ordinary Hall effect (OHE) and the second term denotes AHE. The presence of AHE is considered as one of the strong evidences for intrinsic ferromagnetism in DMSs [43,44]. However, considering the fact that AHE has also been reported in ferromagnetic clusters [45], granular materials [46-48] and inhomogeneous DMS in the hopping transport regime [49], observation of AHE alone cannot support the claim that the DMS under study is a ferromagnet of intrinsic origin, unless it is correlated with ferromagnetism observed by other means and furthermore, secondary phases and precipitates must be absent in the sample.

Before discussing the results, we give a brief description on how the Hall data were collected, processed and used to calculate the carrier concentrations at different temperatures. The major steps involved are as following: (1) the Hall measurements were carried out on Hall bar samples and the raw Hall voltages were measured; (2) the raw data were corrected by subtracting the MR contributions due to electrode mis-alignment, if any; (3) the Hall voltage-field slopes in the linear region for the MR-corrected data at high field are determined for both the positive and negative field region and an average of the two slopes are then used to calculate the carrier concentration using the equation $n = -(I/ed)(B/V)$, where I is the applied current, e is the electron charge, d is the thickness of the sample, B is the applied field and V is the Hall voltage. It should be noted that any influence of magnetic impurities on the Hall voltage at high field region will affect the absolute value of the carrier concentrations obtained in such a procedure. However, it will not affect the validity of discussion below because we are more concerned on the temperature-dependence rather than the absolute values.

In the present case, as expected, $Zn_{1-x}Co_xO$ samples with $x < 0.2$ show only OHE (Fig. 5(a)-(c)). As Co concentration increases, the AHE appears in samples with $x \geq 0.25$. The onset composition at which AHE starts to appear also coincides with the composition at which the Co-rich phase becomes dominant and Co clusters start to appear. All the samples with $x \geq 0.25$ exhibit very clear AHE characteristics, as shown in Fig. 5(d)-(f) (4.2K) and Fig. 5(g)-(I) (300K) for samples F (x=0.25), G (x=0.27) and J (x=0.3), respectively. We now turn back to Fig. 4(f), where the M-H curves obtained by SQUID, AHE and MCD at different energies are compared for sample H. The hysteresis curves obtained by SQUID, Hall and MCD measured at 2.92 eV, are almost identical in shape. However, the MCD curves measured at other photon energies are obviously different from those measured by SQUID and AHE. These results suggest that the dominant phase in this sample is Zn-incorporated CoO which has formed an electrically percolated network with the help of Co clusters.

3.4.2    Magnetoresistance

Besides the Hall effect, magnetoresistance (MR) has also been measured simultaneously at various temperatures in the field range of -6T to 6T (more precisely, perpendicular MR in this case). MR is more sensitive to Co composition than AHE at low doping levels; thus it allows us to study sd interactions in samples without the presence of AHE. The MR curves (Fig. 6(a)-(h)) for samples with $x < 0.25$ are very similar to those reported in literature [50-54]. In Fig. 6(a), we show the MR behaviour of Al-doped ZnO films without Co. A small negative MR is observed, decreasing with temperature, which is characteristic of weak localization [50,51,53]. With doping of Co ($x \leq 0.2$), a positive MR appears at intermediate field values which is superimposed with a negative MR at both low and high applied magnetic field. At low temperature, the field at which the MR changes from positive to negative increases with Co composition but decreases with temperature. Above 10 - 50 K (depending on Co composition), the MR becomes negative in the entire field range for samples A, B, C, D and F (Fig. 6(b)-(e)). The negative MR near zero field exhibits a similar field dependence as that of ZnO:Al; therefore it can be understood as being originated from the destruction of quantum corrections due to weak-localization. With the further increase of magnetic field, the spins of $Co^{2+}$ ions will become increasingly aligned and large s-d interaction will lead to a splitting of the conduction band into spin-up and spin-down sub-bands. The spin splitting of conduction band enhances electron-electron interactions in a disordered system which leads to a positive MR [51,53]. When the field increases further, a negative MR appears due to possibly the increasing alignment of electron spins with those of $Co^{2+}$ ions [51] or the formation of bound magnetic polarons [53]. As the Co composition increases further, the positive MR becomes dominant in a much wider field range and, at x = 0.25, the MR continues to be positive even up to 6T. As it is shown in the inset of Fig. 6(e), starting from x = 0.25, the negative MR peak at low field is no longer a single peak; instead it shows clear hysteresis. The negative MR is relatively insensitive to temperature and becomes dominant over the positive MR above 70 K. For samples with x > 0.25, the dominance of negative MR with hysteresis becomes even more apparent (Fig. 6(f)-(h)) and the positive MR is no longer observable in the entire field region below 6T. In order to focus on the details of MR at low field, in Fig. 6 (f) – (h), we show the MR in the range of -2T to 2T. The samples with x > 0.25 show a typical MR curve for granular-like material at low field superimposed with a slowly changing negative background. The onset Co composition of such MR behaviour again coincides with the Co composition at which the Co-rich phase becomes dominant and Co clusters start to appear. The MR curves of samples with low Co compositions have been analyzed by Dietl *et al* through fitting to theoretical results, which has led to the conclusion that the magnetic properties observed in these samples are due to uncompensated spins either on the surface or inside ZnCoO antiferromagnetic clusters [55].

3.4.3 Differential conductance curve

The above XRD, TEM, SQUID, optical transmittance, MCD, AHE and MR data showed clearly that the presence of strong s-d interaction in diluted $Zn_{1-x}Co_xO$ samples does not lead to the formation of ferromagnetic DMS in our samples; the ferromagnetic signals in concentrated samples are due to secondary phases and precipitates. These observations pose a serious question here: is it after all really possible to obtain intrinsic ferromagnetism in $Zn_{1-x}Co_xO$? The results obtained in this paper alone are insufficient to give a clear answer to this question. However, the analysis on temperature-dependent carrier localization and density may offer some hints. As it has been addressed by many researchers, both electrostatic and magnetic disorder inherently occur in almost all DMS systems. The widely believed scenario of ferromagnetic ordering in highly disordered DMSs is as follows. Unless a ferromagnetic ordering is already established above room temperature, otherwise when the temperature decreases, carriers will become gradually localized at small potential valleys surrounding which BMPs may be formed. The size of the BMP increases with decreasing temperature, which may eventually lead to a ferromagnetic ordering when the BMPs merge globally. However, to our surprise, so far there has been no experimental study on the correlation between carrier localization and magnetic properties in $Zn_{1-x}Co_xO$. The carrier localization and disorder effect can be studied by several different techniques such as scanning tunnelling microscopy, tunnelling through a point contact between a superconductor and a DMS, and differential conductance measurement. We adopted the differential conductance technique because the conductance can be measured using the same Hall bar sample as that for Hall effect and MR measurements and thus it allows us to correlate differential conductance data with those of Hall effect and MR discussed above.

The differential conductance curves, i.e., $dI/dV_{xx}$ versus $V_{xx}$, were measured for all the samples at different temperatures. The results for ZnO:Al and samples A, D, F, J and L are shown in Fig. 7 (a)-(f), respectively. For ZnO:Al, weak localization of carriers at low temperature is reflected in the MR curves shown in Fig. 6(a). In the differential conductance curve, it introduces a "dip" at zero-bias, the so-called zero-bias anomaly (ZBA). The ZBA, in principle, can appear in many different situations. In a 4-point probe measurement configuration like the one used in this study, the influence of sample-electrode contact can be neglected and, apart from weak localization, ZBA is mainly caused by electrostatic potential disorder in the sample. The difference between weak localization and electrostatic potential induced carrier localization can be readily differentiated from the dependence of ZBA on an applied magnetic field. The ZBA for ZnO:Al disappears completely at an applied field of 1T perpendicular to the sample surface, while those for samples with Co doping are insensitive to the external field. The size of ZBA serves as an indicator of carrier localization strength and the shape of the differential conductance curve helps to identify the carrier transport mechanism in individual samples. Although currently there is no theoretical model available to explain quantitatively the shape of differential conductance curves observed here, we do observe clear changes in both the size of ZBA and shape of the conductance curve as the Co composition is varied. As shown in Fig. 7(b) and (c) (x = 0.16 and 0.2), the differential conductance curve for x < 0.2 has roughly a "V" shape, while those for samples with x > 0.2 resembles more a "U" shape, particularly, in the low bias region (Fig. 7 (d)-(f)). The V-shape can be understood as being caused by electrical-field assisted "de-trapping" of carriers localized in shallow potential wells, while the U-shape is attributed to transport across grain boundaries which can be of either a tunnel junction or nanoscale heterojunctions or Schottky junctions [56], depending on the electrical characteristics of the secondary phases. Figure 7(g) shows the dependence of resistivity as well as normalized ZBA (inset) on Co composition at 4.2 K. The latter is defined as $dI/dV_{xx}(2V)/ dI/dV_{xx}(0)$. The sharp increase of both resistivity and ZBA at x = 0.25 agrees well with the finding that secondary phases start to form at this composition. When the Co composition increases further, the secondary phases as well as precipitates increase in density and eventually become electrically percolated in the entire sample, leading to decrease of both resistivity and ZBA. However, comparing to samples at low Co compositions, the dI/dV curve shape is different due to phase separation in these samples. The zero-bias resistivity, as shown in Fig. 7(h), shows all samples at low and high compositions exhibiting a typical "dirty metal" behaviour, while those in-between behave like an insulator due to strong disorder caused by onset of phase separation.

3.4.4 Carrier concentration dependence on temperature

We finally turn to the temperature-dependence of carrier concentration in samples with different Co compositions, as shown in Fig. 8(a)-(h). Also shown in the insets are the normalized ZBA, as defined above, at different temperatures. The large fluctuation of carrier concentration for low-resistivity samples are caused by the small OHE signal. However, it can be seen very clearly that carrier localization indeed occurs at low temperature, in particular in samples with Co composition below 0.25. The high-temperature over low-temperature carrier density ratio agrees well with the same ratio of ZBA for different samples. This suggests strongly that carriers are localized in potential valleys at low temperature and become de-trapped as temperature increases. Now the question is: are the carriers localized in Co-rich or less regions? The answer to this question is important because it will determine if and how the carriers will affect magnetic properties of the samples. Structural and optical analyses showed that Co is soluble in ZnO for x < 0.2. However, when the Co composition reaches a few percents, it is inevitable that Co-rich regions will form in a background of less Co regions. As it is revealed by optical transmission measurement, the bandgap of $Zn_{1-x}Co_xO$ in Co-less regions is larger than that in Co-rich regions. Therefore, the electrons will get localized in Co-rich regions at low temperatures. If carrier mediated ferromagnetic ordering takes place in the Co-rich regions before secondary phases or precipitate form, one will be able to obtain intrinsic DMS. However, the results obtained in this study showed otherwise, i.e., carrier-mediated ferromagnetic ordering didn't take place before the secondary phases or precipitate form which results in correlated ferromagnetism of extrinsic origin at high Co compositions. As it was shown recently by Venkatesan *et al.* the doping of Al can also affect the Co distribution and formation of Co clusters in ZnCoO [57]. Therefore, it is quite challenging to establish the optimum doping window for both magnetic and non-magnetic impurities. One of possible solutions is to introduce dopant which supplies both unpaired spins and free carriers [58].

Summarising the results from structural, magnetic and electrical transport studies, the samples studied can be divided into following different regimes:

1) x < 0.2: In this regime, Co is soluble in ZnO and strong s-d interaction is observed in the MR measurement; however, there is no AHE observed. The hysteresis observed by SQUID is likely due to uncompensated spins on the surface or inside the ZnCoO anti-ferromagentic clusters [55].

2)    $x = 0.2 - 0.25$: Onset of secondary phase formation occurs in this region, though Co still exists dominantly in the 2+ valence state. Columnar structures start to form which eventually become electrically percolated networks of Zn-incorporated CoO. The ferromagnetism increases rapidly with increasing the Co composition due to the formation of $ZnCo_2O_4$ and Co clusters.

3)    $x \geq 0.25$: The sample is inhomogeneous and consists of Co-incorporated ZnO, Zn-incorporated CoO, $ZnCo_2O_4$ and Co clusters. The formation of electrically percolated ferromagnetic networks leads to the observation of AHE.

Although the mechanism for the formation of clusters or columnar structures is not well understood at present, one of the possible mechanisms might be due to the spinodal nano-decomposition induced by strong attractive interactions between the magnetic impurities [59,60]. According to this model, due to the formation of magnetic nano-clusters induced by spinodal decomposition, the super-paramagnetic blocking temperature of the DMS is enhanced and hysteretic magnetic response can be observed at finite temperature even if Curie temperature of the whole system is nearly zero. This may explains why we observed FM-behaviour by SQUID but not in electrical measurements.

## 4    Summary

In summary, we have investigated systematically the structural, optical, electrical and magnetic properties of $Zn_{1-x}Co_xO$:Al prepared by sputtering. We have shown that, although Co is soluble in $Zn_{1-x}Co_xO$ for $x < 0.2$, Co-rich regions begin to form even at a few percent of Co doping accompanied by carrier localization at low temperature in the same regions. However, these Co-rich regions eventually develop into secondary phases/precipitates with a further increase of Co doping before carrier-mediated ferromagnetic ordering takes place. For the samples investigated, the Co-rich regions become electrically percolated at about x = 0.25 beyond which ferromagnetic properties have been observed not only in magnetometer and magnetooptical measurements but also in Hall effect, due to formation of extrinsic ferromagnetic networks. Although strong s-d interaction and carrier localization have been observed in samples with x < 0.2, the dominant phase of these samples is paramagnetic with a in-plane anisotropy. The hysteresis observed by SQUID is presumably due to uncompensated spins on the surface or inside the ZnCoO antiferromagentic clusters. If the theoretical predictions are correct, i.e., high carrier concentrations and Co doping are necessary for realizing carrier-mediated DMS, then strategies must be found to control the carrier concentration independently as well as to further enhance the interactions between carriers and magnetic impurities before one can obtain true carrier-mediated DMS in the ZnO-based material systems.

## Acknowledgements

The authors are grateful to Koji Ando for the MCD measurement and detailed analysis of measurement results and JZ Wang for help with the cryostat measurements. The authors are also grateful to T. Andrearczyk and T. Dietl for their help on the interpretation of the MR results. The work at the National University of Singapore was supported by the A*-STAR under Grant No. R-398-000-020-305. The work at the University of Michigan was supported by the U.S. National Science Foundation (NSF) under Grant No. DMR 0308012.

**Table**

Table 1. List of samples used in this study.

| Sample Name | Co Sputtering Power (W) | Co composition (%) |
|---|---|---|
| A | 3 | 4.60 - $Zn_{0.95}Co_{0.05}O$ |
| B | 8 | 13.70 - $Zn_{0.86}Co_{0.14}O$ |
| C | 10 | 15.90 - $Zn_{0.84}Co_{0.16}O$ |
| D | 15 | 19.80 - $Zn_{0.80}Co_{0.20}O$ |
| E | 20 | 23.53 - $Zn_{0.76}Co_{0.24}O$ |
| F | 25 | 24.43 - $Zn_{0.75}Co_{0.25}O$ |
| G | 30 | 27.30 - $Zn_{0.73}Co_{0.27}O$ |
| H | 32 | 28.70 - $Zn_{0.71}Co_{0.29}O$ |
| I | 35 | 30.03 - $Zn_{0.70}Co_{0.30}O$ |
| J | 40 | 30.43 - $Zn_{0.70}Co_{0.30}O$ |
| K | 45 | 29.60 - $Zn_{0.70}Co_{0.30}O$ |
| L | 50 | 33.40 - $Zn_{0.67}Co_{0.33}O$ |

**Figure Captions**

Fig. 1.  X-ray diffraction patterns of samples (a) C, D, (b) E, F, (c) I and L grown on $Al_2O_3$ (001) substrates, with peak positions of ZnO, $Co_3O_4$, Co, $ZnCo_2O_4$ and $Al_2O_3$ indicated.

Fig. 2.  Low magnification TEM images of samples (a) D, (b) E and (c) H. Insets are the corresponding electron diffraction patterns. (d) Electron diffraction pattern of sample H. (e) HRTEM image of a selected region containing Co clusters.

Fig. 3.  (a) Optical transmission spectra of Al-doped ZnO and samples C, D, E, F, H, J and L, '*' marks the energy levels used to determine hysteresis curves shown in Figs. 3(c) & (d). (b) Differential transmission spectra with respect to wavelength. The inset shows the wavelength of both the blue (A) and red (B) shifted peaks as a function of the Co composition. The solid lines are fitted to the equation of Qiu *et al*.[35] for peak B and the result of Reference [33] for peak A. (c) and (d) MCD curves of sample H at 6 K and 300 K, respectively, for wavelengths at 325 nm, 343 nm, 425 nm and 756 nm.

Fig. 4.  (a) In-plane M-H curves of samples A, B, D and H. (b) Saturation magnetization as a function of Co composition at 300 K. (c) Coercivity as a function of Co composition at different temperatures. (d) M-H curves (both in plane and perpendicular to sample surface) at 300 K for sample C. (e) M-H curves (both in plane and perpendicular to sample surface) at 10K and 300 K for sample H. (f) Comparison of hysteresis curves of sample H determined by MCD at different energies with those obtained by SQUID and AHE.

Fig. 5. Hall voltage as a function of applied magnetic field for (a) Al-doped ZnO and samples A - (b), D - (c), F - (d), G - (e) and I - (f) at 4.2K and F - (g), G - (h) and I - (i) at 300K. For Al-doped ZnO and samples A and D, the Hall measurements have been carried out in three different configurations, as indicated by the legends in the figures. Note that the MR contribution due to mis-alignment of the contact electrodes has been subtracted out from the measured Hall voltage. However, the nonmagnetic field dependent offset is intentionally not corrected so as to reveal the true Hall response of the sample to external magnetic field.

Fig. 6. MR curves of (a) Al-doped ZnO and samples A - (b), B - (c), D - (d), F - (e), G - (f), J - (g) and L - (h) at various temperatures as a function of applied magnetic field. Inset of (e) shows the MR curve for sample F at 50 K in the field range of -1 T to 1 T.

Fig. 7. Differential conductance curves at various temperatures as a function of applied voltage for Al-doped ZnO (a) and samples A - (b), D - (c), F - (d), J - (e) and L - (f). (g) Resistivity versus Co composition at 300 K, with the inset showing zero bias anomaly dependence on Co composition at 4.2 K. (h) Resistivity versus temperature for various samples (inset: Al-doped ZnO).

Fig. 8. Carrier concentration of (a) Al-doped ZnO and samples A - (b), B - (c), D - (d), F - (e), G - (f), J - (g) and L - (h) as a function of temperature. Inset of each graph shows zero bias anomaly (ZBA) as a function of temperature for the corresponding sample. The Hall voltage that has been used to calculate the carrier concentration was obtained as following: $V_{xy} = [V_{xy} (B_{max}) - V_{xy} (-B_{max})]/2$, where $B_{max}$ is the maximum field used for the Hall effect measurement.

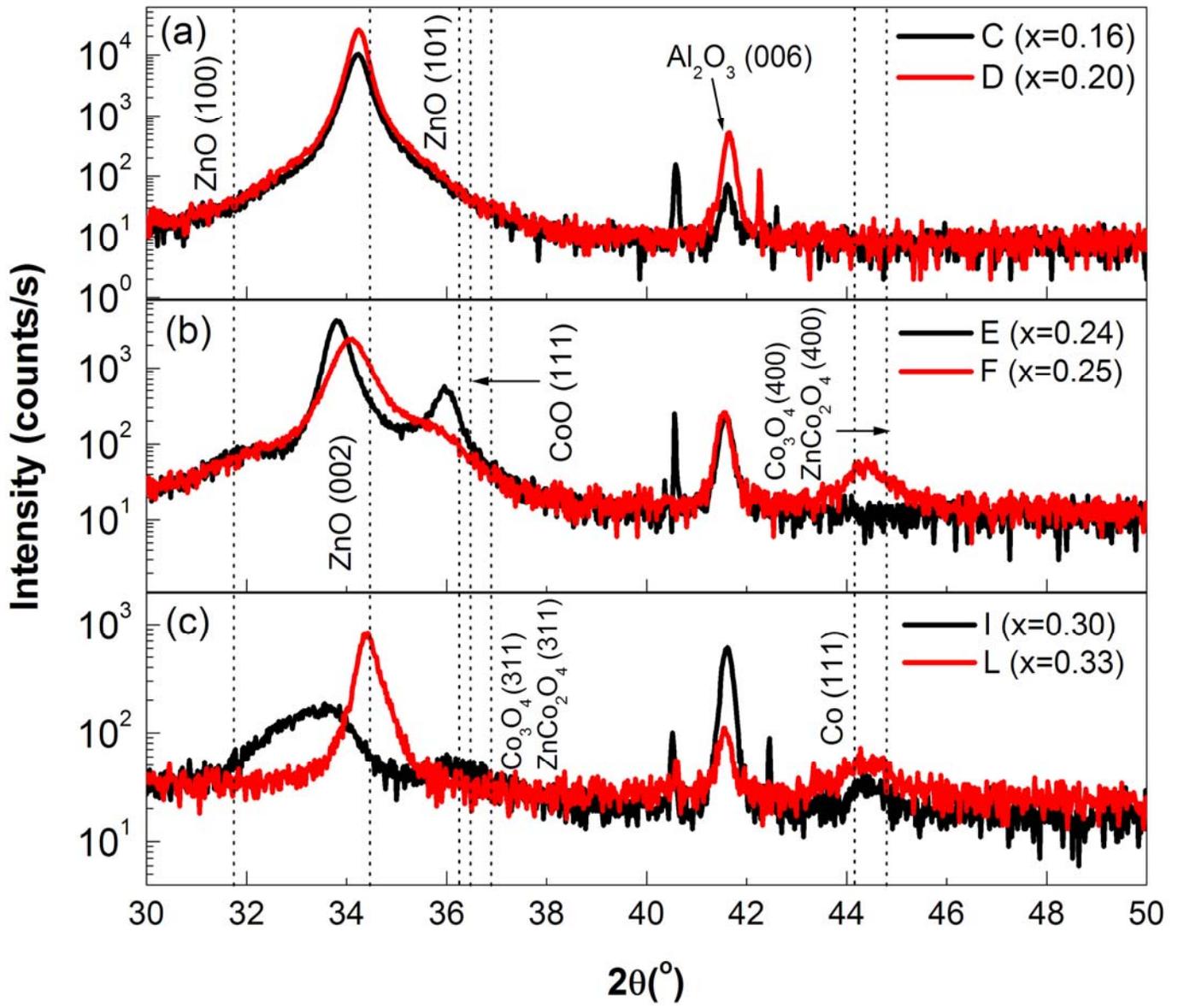

Fig.1
M. Tay et al.

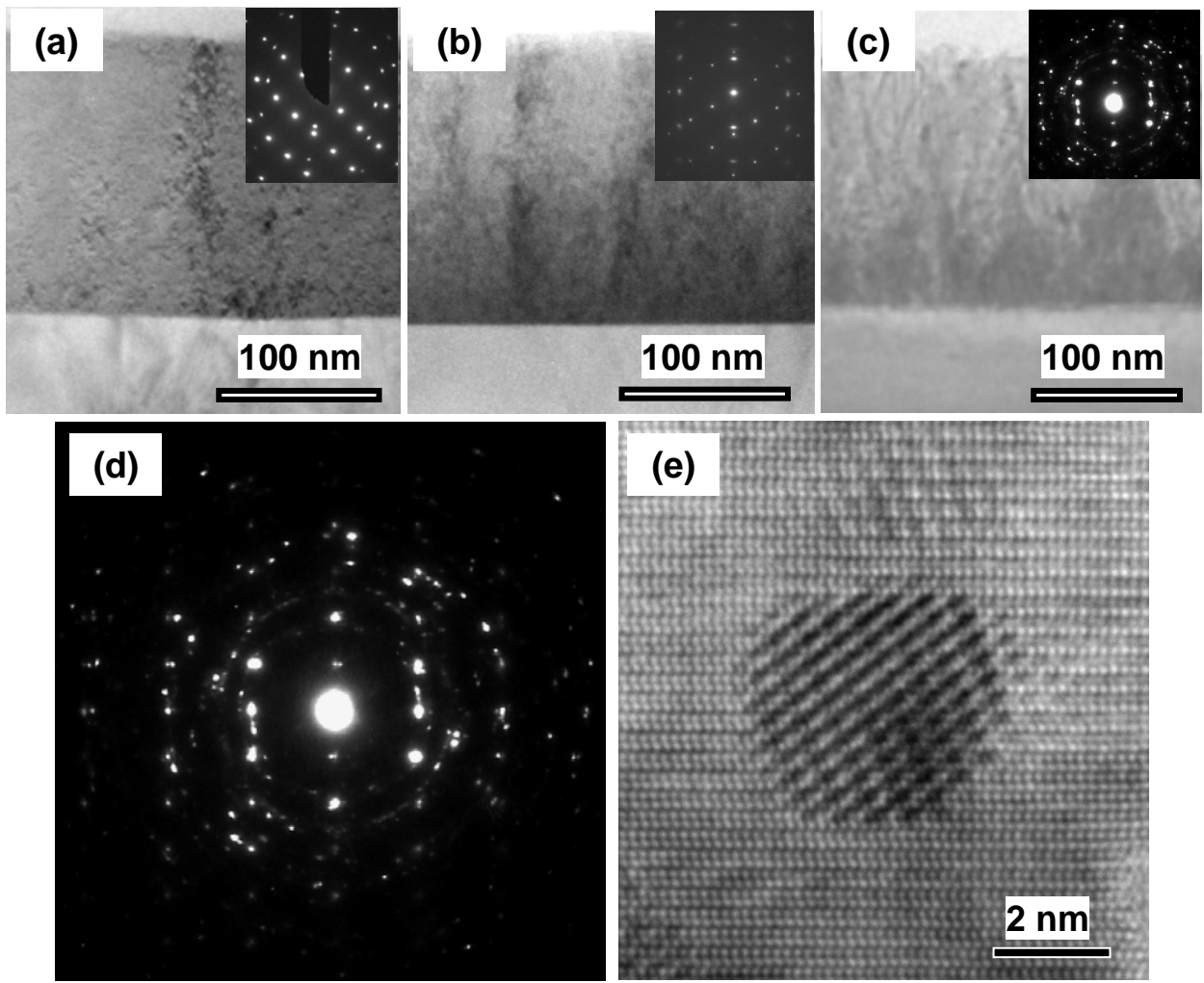

Fig.2
M. Tay et al.

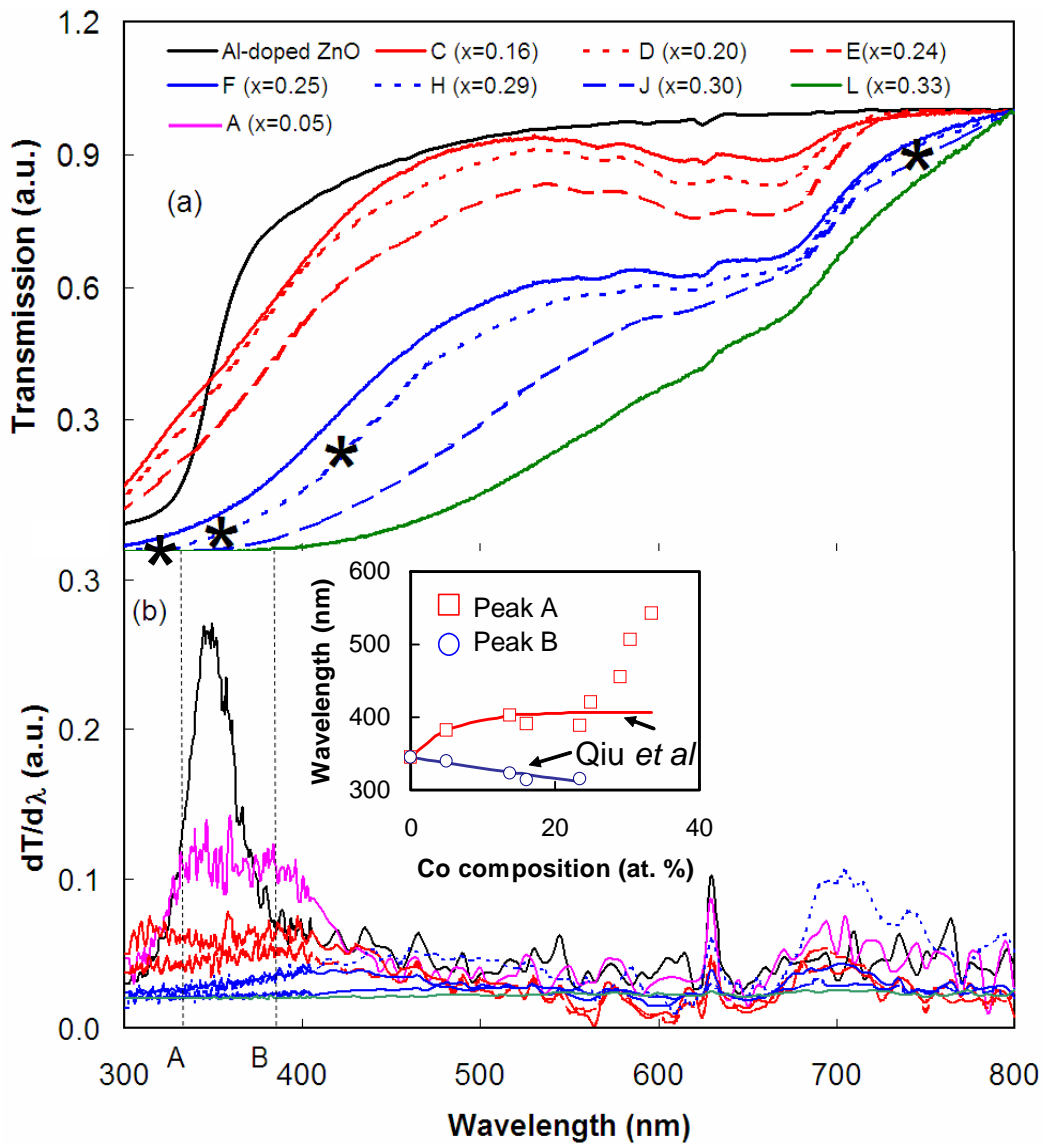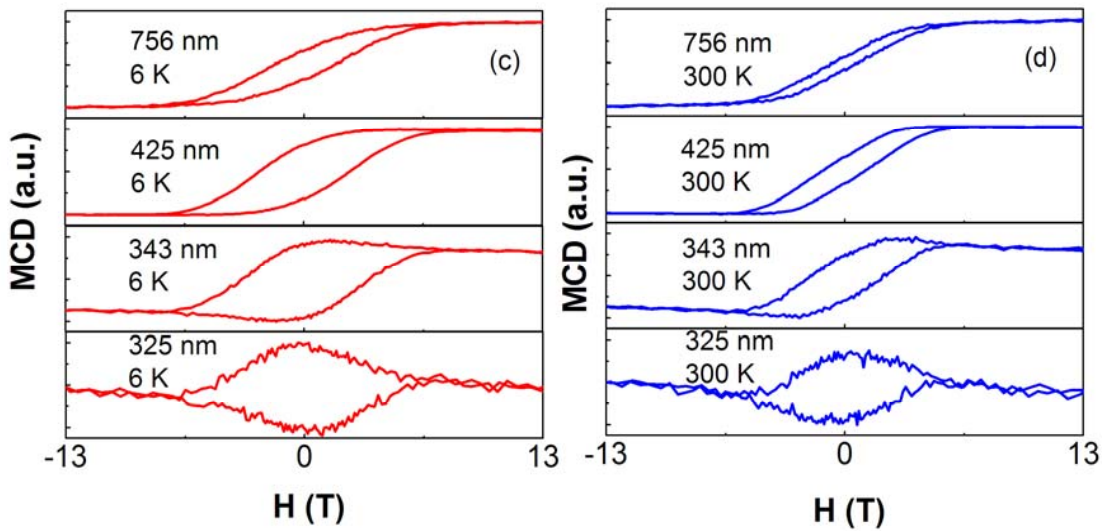

Fig.3
M. Tay et al.

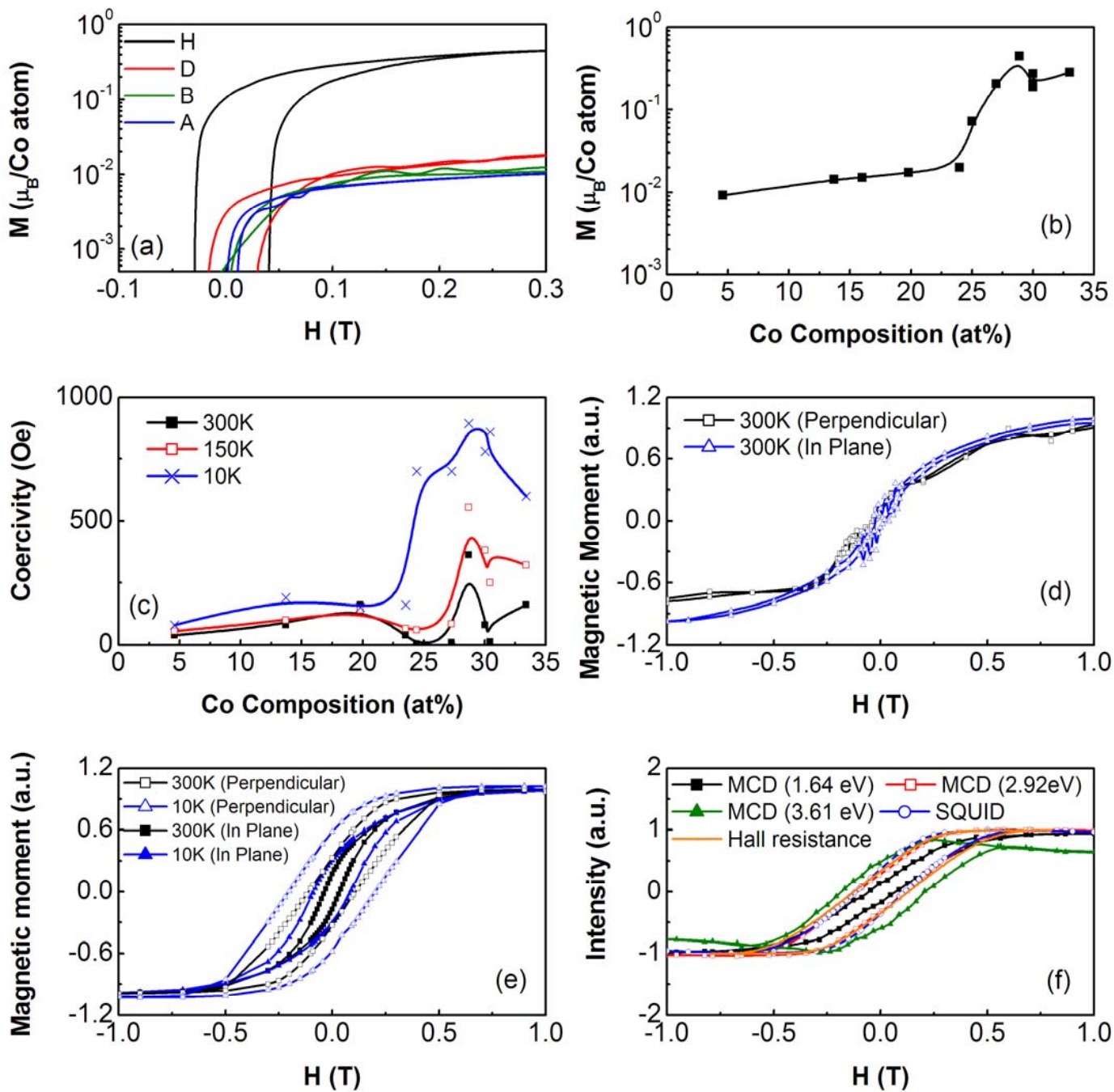

Fig.4
M. Tay et al.

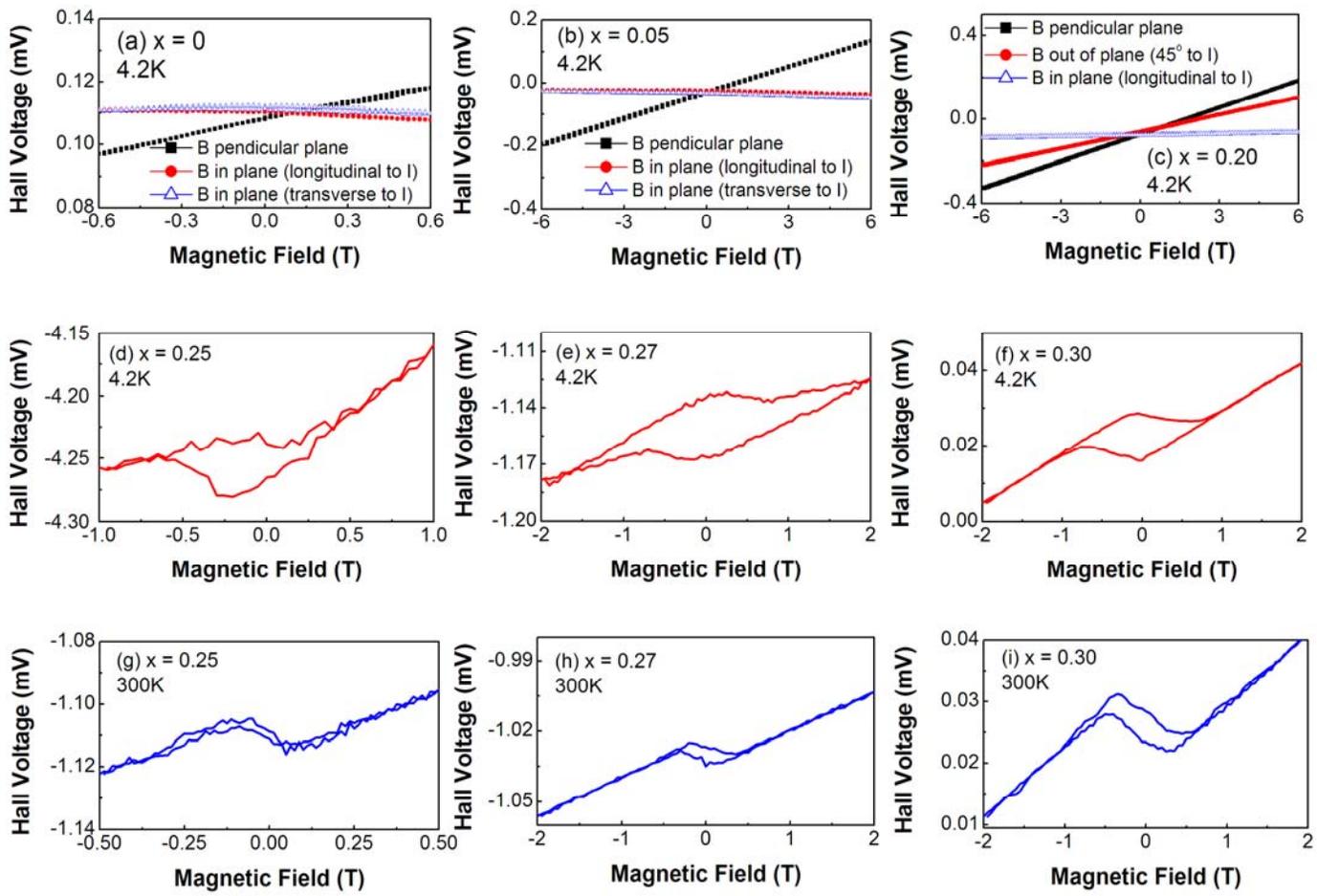

Fig.5
M. Tay et al.

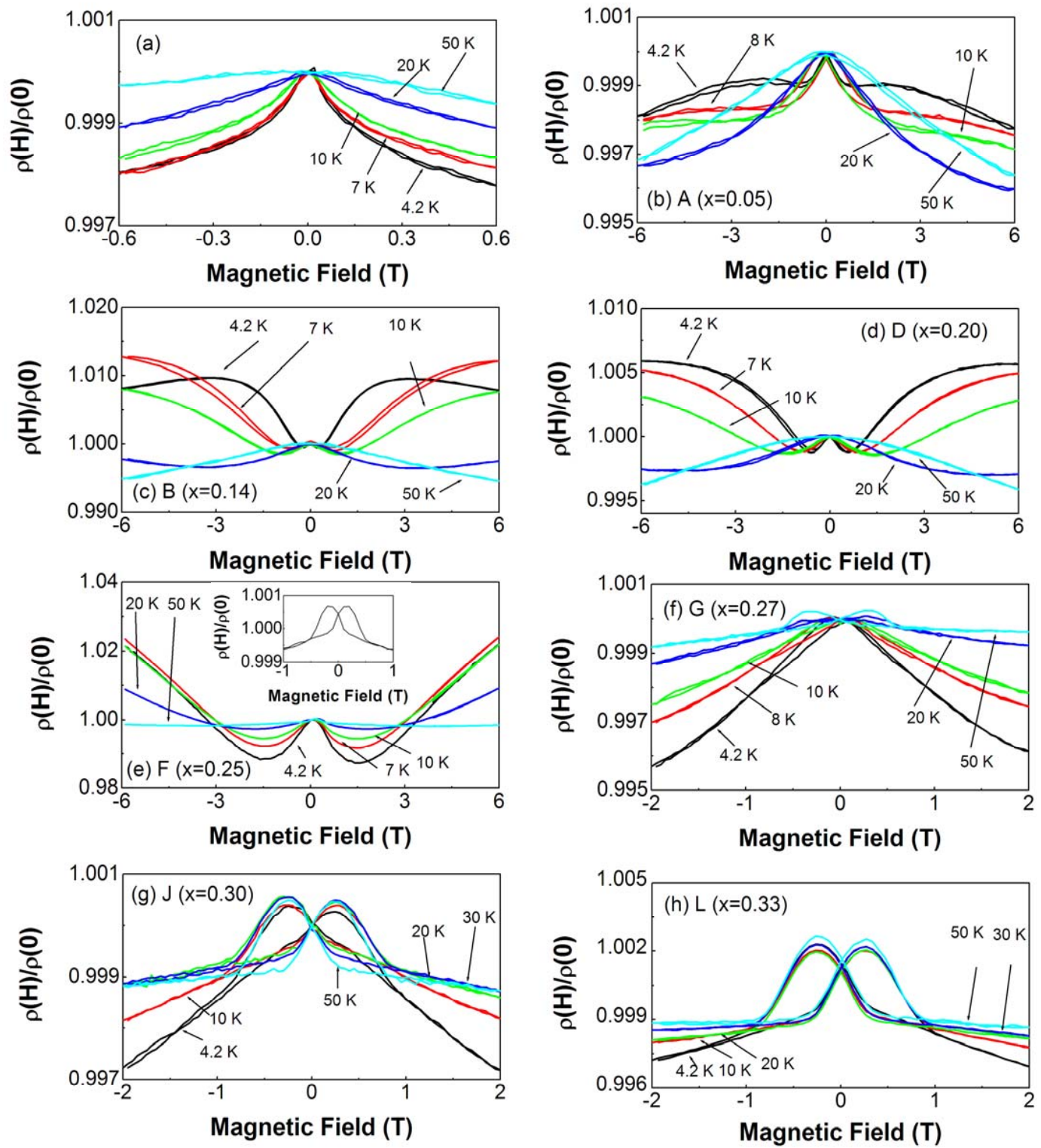

Fig.6
M. Tay et al.

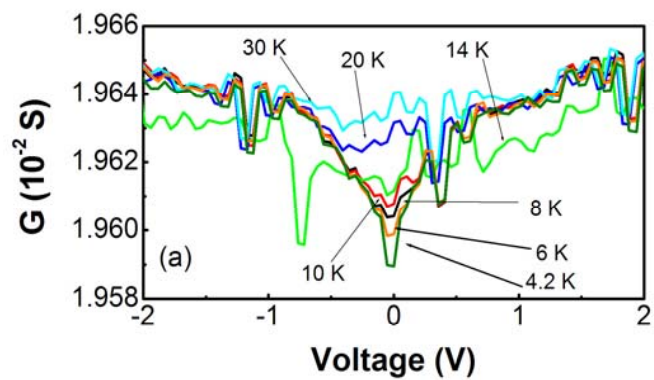
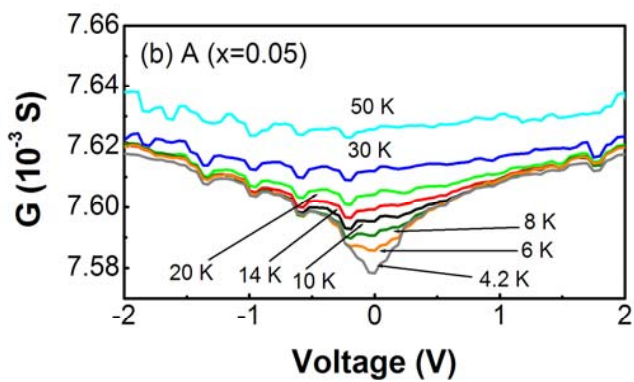
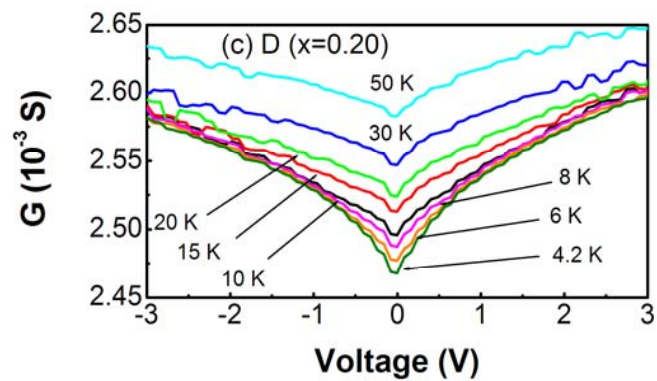
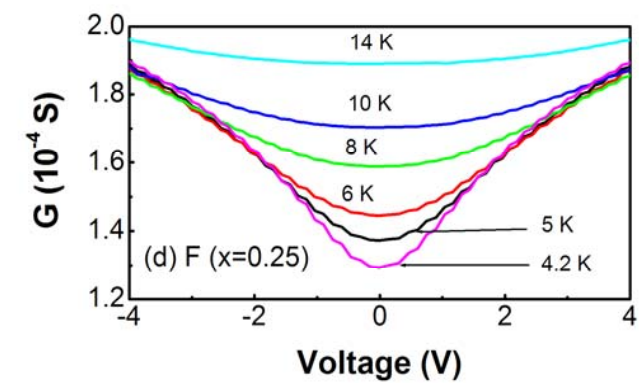
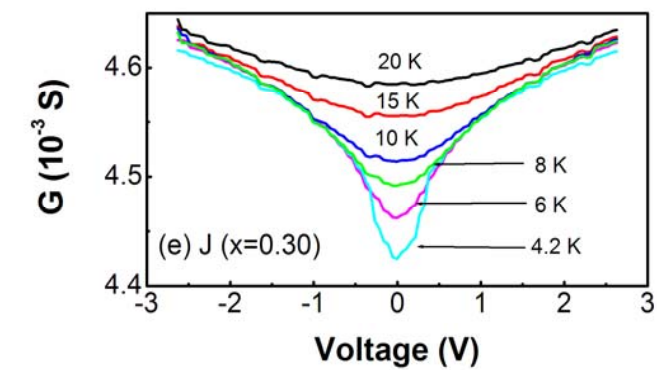
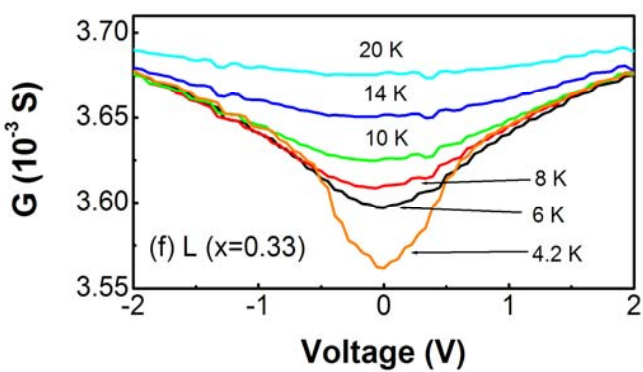
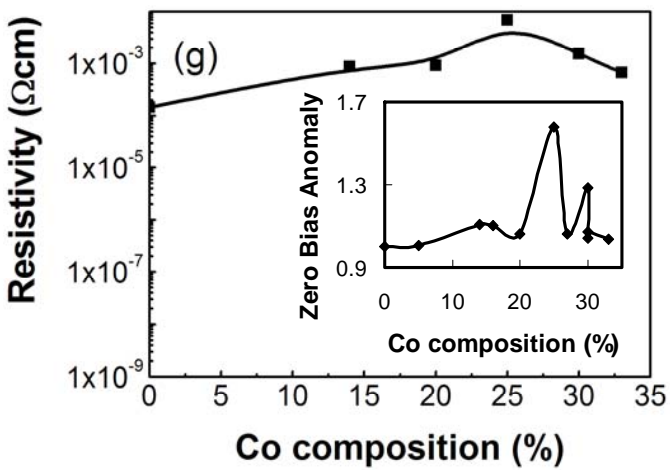
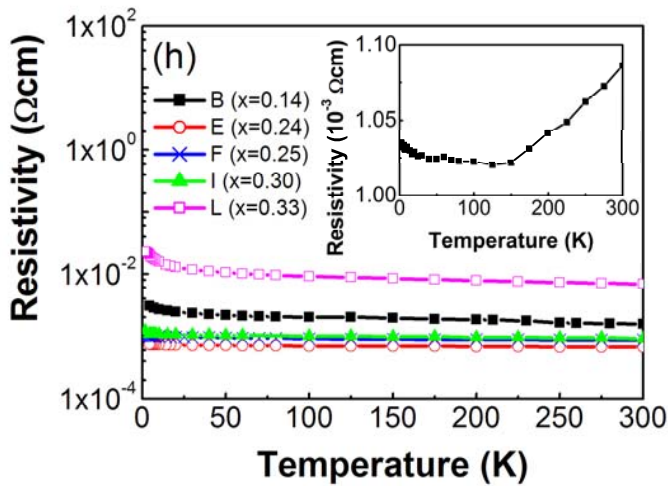

Fig.7
M. Tay et al.

Fig.8
M. Tay et al.